# Unitary Scheme Model Calculations of the Ground- and Excited-State Characteristics of $^3$H and $^4$He

S. B. Doma[1)] and H. S. El-Gendy[2)]

[1)] Faculty of Science, Alexandria University, Alexandria, Egypt
sbdoma@alexu.edu.eg
[2)] Faculty of Science & Art, Shaqra University, Shaqra, KSA
hselgendy1@yahoo.com

**Abstract**
The ground and excited states of the $^3$H and $^4$He nuclei are studied in the framework of group-theoretical methods. Basis functions of the unitary scheme model corresponding to even numbers of quanta of excitation in the range $0 \leq N \leq 20$ are constructed for the even-parity states of these two nuclei, and bases with $1 \leq N \leq 19$ are constructed for the odd-parity states of the $^4$He nucleus. The ground-state and first-excited-state energies and wave functions, the ground S-, P- and D-state probabilities, the root-mean-square radius and the magnetic dipole moment of triton are calculated. Furthermore, for the $^4$He nucleus, the spectrum and the wave functions, the ground S-, P- and D-state probabilities, the root-mean-square radius and the total integral cross section of the dipole electric transition accompanying the photoabsorption of $\gamma$ quanta by this nucleus are calculated. The GPT and AV8' two-body interactions and the UIX three-nucleon interaction are used in these investigations. Moreover, the convergence of the calculations is examined by incrementally extrapolating the nuclear characteristics calculated for $N \leq 20$ to $N = 30$ for both nuclei.

## 1. Introduction

The nuclear model that can be applied to a given nucleus depends on the knowledge that is available from the experimental data regarding that nucleus and on a specific mathematical description of the Hamiltonian operator that considers all possible motions of the nucleons. The purpose of a nuclear model is to provide a theoretical description that allows one to evaluate the different characteristics of nuclei to within fairly good agreement with the corresponding experimental data. The success of such a model is judged by its ability to describe all of the ground- and excited-state characteristics of nuclei with a sufficiently small number of parameters in the nuclear Hamiltonian. The assumption of the independent motion of a nucleon inside a nucleus in an average potential created through the action of the other nucleons on that nucleon, to a zeroth-order approximation, has achieved great success in studies of nuclear structure. This can be seen for either deformed or nondeformed nuclei by means of an average anisotropic potential or an isotropic oscillator potential, respectively. Group-theoretical methods play an important role, especially in describing the bound states of light nuclei and in calculating the different characteristics of these nuclei. These methods provide us with accurate techniques for calculating the different matrix elements of the operators associated with nuclear characteristics, as can be seen from the success of the nuclear shell model [1] and the unitary scheme model (USM) [2-4]. These methods provide us with a complete description of the structure of a nucleus based on the shape of its Hamiltonian operator and the commutation relations satisfied by its constituents. They offer a direct approach for constructing the complete nuclear wave functions of the different states of a given nucleus. Once these nuclear wave functions have been constructed, it is easy to calculate the different characteristics of the nuclear states. Group-theoretical methods of expanding the total antisymmetric nuclear wave function in terms of a complete set of orthonormal functions, i.e., basis functions, have been extensively used, especially for nuclei with $3 \leq A \leq 6$ [2]. The USM, also known as the translationally invariant shell model (TISM) [2-4], has yielded good results for the structure of light nuclei with $A \leq 7$ through the use of nucleon-nucleon interactions [5-10]. The model was given its name due to the major role played by the unitary group in the construction of the nuclear wave functions. In the USM, a nucleus is regarded as a system of non-interacting quasi-particles; this treatment enables us to apply algebraic methods to study the general features of the matrix elements of operators that correspond to physical quantities. The USM is based on group-theoretical methods of classifying the basis functions. The basis functions of this model are constructed such that they will have a certain symmetry with respect to the interchange of particles and will have a definite total angular momentum $J$ and isotopic spin $T$. The basis functions of the USM for a nucleus with mass number $A$ are then expanded in the form of products of two types of functions, one corresponding to the set of $A'$ nucleons and the other corresponding to the set of $A''$ nucleons, where $A' + A'' = A$, by means of many-particle fractional parentage coefficients (FPCs) [2,5,6,10]. Using two- and three-body operators, it is then possible to calculate the corresponding matrix elements with respect to the different nuclear states by means of two- and three-particle FPCs. The expectation values of on-body operators can also be used by means of one-particle FPCs. In principle, the calculated results for the nuclear characteristics should be independent of the particular bases chosen when the number of terms in the expansion is sufficiently large. The inclusion of all possible bases in the expansion is too difficult since the matrices of the two-particle FPCs corresponding to these bases rapidly grow in size. Therefore, it is fundamentally important to identify some rules that will allow us to reduce the number of bases used. Some such rules have been adopted for light nuclei in ref. [2] and for the $^6$Li nucleus in ref. [6]. On the other hand, recent experimental results for three-body systems have unambiguously shown that calculations based only on nucleon–nucleon forces fail to accurately describe many experimental observables; instead, one needs to include effects that extend beyond the realm of two-body potentials. In addition, microscopic calculations for light nuclei and nuclear matter [11] have indicated that it is difficult to explain the observed binding energies and densities if we assume a non-relativistic nuclear Hamiltonian containing only two-nucleon interactions, consistent with the nucleon-nucleon scattering data at low energies ($E_{lab} < \sim 400 \, MeV$).

Since nucleons are composite objects made up of quarks and gluons, we cannot approximate their interactions through a sum of two-body terms alone. The mesonic degrees of freedom can also generate potentials involving three bodies and more in the Hamiltonian, in which only the nucleon degrees of freedom are retained. However, since the energies obtained with a Hamiltonian that contains only two-body potentials are not far from those observed in experiments, we expect that the contribution of the many-body potentials should be small compared with that of the two-body interactions in the realm of nuclear physics, and indeed, only three-body potentials may be important.

Marsden et al. [12] studied the $^3$H and $^3$He nuclei with a realistic nucleon-nucleon potential and the Tucson-Melbourne (TM) three-body interaction using a translationally invariant harmonic oscillator (HO) basis. They replaced the nucleon-nucleon potential with the no-core shell model two-body effective interaction, and the three-nucleon interaction was added without any renormalization. The authors also studied the convergence of the approach by changing the basis size. Furthermore, they examined the dependence of the binding energies on the TM cut-off parameter L. The results showed promise for the construction of three-body effective interactions, including a three-nucleon interaction, for use in future *ab initio* no-core shell model nuclear structure calculations for $A > 3$ systems.

The effects of three-nucleon interactions have been recently studied in ref. [13], where the effects of different three-nucleon interactions in $p$-$^3$He elastic scattering at low energies were calculated for four-nucleon scattering observables by using the Kohn variational principle and the technique of hyperspherical harmonics. Meanwhile, the effects of two- and three-body hyperon-nucleon interactions in $\Lambda$ hypernuclei have been studied by assessing the relative importance of the two- and three-body hyperon-nucleon forces and by studying the effect of the hyperon-nucleon-nucleon interaction in closed-shell $\Lambda$ hypernuclei for A = 5 to 91 [14,15]. Moreover, Cipollone et al. [16] extended the formalism of self-consistent Green's function theory to include three-body interactions and applied it to isotopic chains around oxygen for the first time. Furthermore, Wiringa et al. [17] used the realistic Argonne $v_{18}$ two-nucleon potential and Urbana three-nucleon potentials to generate accurate variational Monte Carlo (VMC) wave functions for the $A \leq 12$ nuclei.

The ab initio no-core shell model (NCSM) is a well-established theoretical framework with the aim of providing an exact description of nuclear structure starting from highly precise descriptions of the interactions between nucleons. Barrett, Navrátil, and Vary [18] discussed, in detail, the extension of the ab initio NCSM to nuclear reactions and outlined some of the promising future directions for research emerging from the foundation provided by the NCSM, including a microscopic non-perturbative framework for a theory with a core. In the NCSM, Forssén, Navrátil and Quaglioni [19] considered a system of point-like non-relativistic nucleons that interact through realistic inter-nucleon interactions. They considered two-nucleon interactions that reproduce nucleon-nucleon phase shifts with high precision, typically up to 350 MeV in lap energy. They also included three-nucleon interactions with terms related to, e.g., two-pion exchanges with an intermediate delta excitation. Both semi-phenomenological potentials (based on meson-exchange models) and modern chiral interactions were considered.

Calculations within a three-body translationally invariant HO basis using realistic two- and three-nucleon forces have previously been performed for the three-nucleon system (see, e.g., [12]) and the four-nucleon system (see, e.g., [20,21]). These calculations were based on the translationally invariant form of the NCSM, which is equivalent to the TISM except that the antisymmetrization of the wave function is achieved not by means of group theory but rather by diagonalizing the antisymmetrization operator and retaining the antisymmetric eigenstates as basis states.

Three- and four-nucleon systems have also been prominently studied by means of numerically exact few-body approaches (such as the Faddeev, Faddeev-Yakubovsky and hyperspherical harmonics approaches starting from realistic two- and three-nucleon forces; see, for example, [22-24]). A review can be found in [25].

In previous papers, Doma et al. [10] applied the USM with a number of quanta of excitation satisfying $0 \leq N \leq 8$ to investigate the ground-state wave function, the binding energy, the first-excited-state energy and the root-mean-square radius of $^3$H by using the Gogny, Pires and de Tourreil (GPT) potential [26], the Hu and Massey (HM) potential [27] and an effective potential proposed by Vanagas [2]. Furthermore, for the

$^4$He nucleus, Doma [28] investigated the binding energy, the structure of the wave functions, the excitation spectrum, the root-mean-square radius and the total integral cross section of the photoabsorption of $\gamma$ quanta by this nucleus by using the USM with basis functions corresponding to $N \leq 8$ and the GPT potential. Moreover, Doma et al. [6] investigated the binding energy, the structure of the wave functions, the excitation spectrum, the root-mean-square radius, the magnetic dipole moment, the $ft$ value for the allowed $\beta^-$ transition, and the partial and integral cross sections of the dipole electric transition accompanying $\gamma$ absorption for the $A = 6$ nuclei by using the USM with the GPT and HM potentials. It was concluded in the cited paper that the GPT potential, which was the most suitable potential for the nucleon-nucleon interaction at small distances, yielded the best fits to the calculated characteristics of the $^6$Li nucleus and that the agreement between the theory and the experimental data was considerably improved by including higher configurations in the analysis.

Accurate solutions for the distribution of the HO excitations have been calculated for $^4$He by W. Horiuchi and Y. Suzuki [29]. They developed a method for calculating the occupation probability of the number of HO quanta for a precise few-body wave function obtained in a correlated Gaussian basis. They used different nucleon-nucleon interactions in these calculations. Both the tails of the wave functions and the cluster configurations are well accounted for in these calculations. Thus, the resulting distribution is expected to be realistic. Tensor correlations and short-range correlations both play a crucial role in enhancing the probability of high HO excitations. For the excited states of $^4$He, the interaction dependence is much less because high HO quanta are mainly necessary for describing the relative-motion function between the $^3$H + $p$ and $^3$He + $n$ clusters.

In the present study, we applied the USM with basis functions corresponding to even numbers of quanta of excitation in the range $0 \leq N \leq 20$ to investigate the ground-state and first-excited-state wave functions and energies, the S-, P- and D-state probabilities $(P_S, P_P, P_D)$, the root-mean-square radius, and the magnetic dipole moment of triton. In addition, for the $^4$He nucleus, the basis functions of the USM with numbers of quanta of excitation in the range $0 \leq N \leq 20$ were used to calculate the wave functions and energies of the ground state and the even- and odd-parity excited states; the S-, P- and D-state probabilities; the root-mean-square radius; and the total integral cross section of the dipole electric transition accompanying the photoabsorption of $\gamma$ quanta by this nucleus. To perform these calculations, we used two nucleon-nucleon interactions and a three-nucleon interaction. The first nucleon-nucleon interaction is the GPT potential [26], which is a smooth, realistic local nucleon-nucleon force. It fits the two-body data and the deuteron binding energy, quadrupole moments and magnetic moments and is suitable for nuclear Hartree-Fock calculations. The second is the well-known AV8' nucleon-nucleon interaction [30]. For the three-body interaction, we used the Urbana IX model (UIX) interaction [31]. Moreover, the convergence behaviour of the USM calculations was examined by incrementally extrapolating the results for $N \leq 20$ to $N = 30$ for both nuclei.

**2. The Hamiltonian and the Total Nuclear Wave Function**

The Hamiltonian $\mathcal{H}$ of a nucleus consisting of $A$ nucleons, interacting via two-body potentials, can be written in terms of the relative coordinates of the nucleons in the following form [28]:

$$\mathcal{H} = \frac{1}{2m} \sum_{i=1}^{A} \boldsymbol{p}_i^2 + \frac{1}{2} \sum_{1=i \neq j}^{A} \sum_{j=1}^{A} V(|\boldsymbol{r}_i - \boldsymbol{r}_j|). \qquad (2.1)$$

The translational invariance of the Hamiltonian $\mathcal{H}$ permits the separation of the centre-of-mass motion, and consequently, the Hamiltonian corresponding to the internal motion becomes

$$H = \mathcal{H} - \frac{1}{2mA}\left(\sum_{i=1}^{A} \boldsymbol{p}_i\right)^2. \tag{2.2}$$

By adding and subtracting an oscillator potential referred to the centre of mass, the internal Hamiltonian $H$ can be rewritten in terms of the relative coordinates of the nucleons in the form

$$H = H^{(0)} + V', \tag{2.3}$$

where

$$H^{(0)} = \frac{1}{A}\sum_{1=i<j}^{A}\left[\frac{1}{2m}(\boldsymbol{p}_i - \boldsymbol{p}_j)^2 + \frac{1}{2}m\omega^2(\boldsymbol{r}_i - \boldsymbol{r}_j)^2\right] \tag{2.4}$$

is the well-known USM Hamiltonian, also known as the TISM Hamiltonian, and

$$V' = \sum_{1=i<j}^{A}\left[V(|\boldsymbol{r}_i - \boldsymbol{r}_j|) - \frac{m\omega^2}{2A}(\boldsymbol{r}_i - \boldsymbol{r}_j)^2\right] \tag{2.5}$$

is the residual two-body interaction.

The energy eigenfunctions and eigenvalues of the Hamiltonian $H^{(0)}$ are given by [28]

$$|A\,\Gamma; M_L M_S T M_T\rangle \equiv |A\,N\{\rho\}(\nu)\alpha[f]LS; M_L M_S T M_T\rangle, \tag{2.6}$$

$$E_N^{(0)} = \left\{N + \frac{3}{2}(A-1)\right\}\hbar\omega. \tag{2.7}$$

The functions expressed in (2.6) form a complete set of functions, or bases. It is easy to construct bases that have a definite total momentum $J$ in the following form [8,9,28]:

$$|A\,\Gamma J M_J T M_T\rangle = \sum_{M_L+M_S=M_J}(LM_L, SM_S|JM_J)|A\,\Gamma; M_L M_S T M_T\rangle, \tag{2.8}$$

where the $(LM_L, SM_S|JM_J)$ are the Clebsch-Gordan coefficients of the rotational group $SO_3$. The nuclear wave function of a state with total momentum $J$, isospin $T$, and parity $\pi$ can be constructed as follows [8,9,28]:

$$|A\,J^\pi T M_J M_T\rangle = \sum_{\Gamma} C_{\Gamma}^{J^\pi T}|A\,\Gamma J M_J T M_T\rangle, \tag{2.9}$$

where the $C_{\Gamma}^{J^\pi T}$ are the state-expansion coefficients. In the sum on the right-hand side of (2.9), the number of quanta of excitation $N$ can be either an even or odd integer depending on the parity of the state $\pi$. It is obvious that the USM Hamiltonian is free of spurious states. The spurious states that must be eliminated correspond to the non-zero motion of the centre of mass of the entire nucleus.

It is well known that three-body forces are important for describing the properties of finite nuclei. The parameters in the nucleon-nucleon potential may not be unique, or there may be some redundant parameters necessary to reproduce the deuteron properties. To investigate these possibilities, we consider the following Hamiltonian operator, which includes three-body forces:

$$H = H^{(0)} + V' + V'', \tag{2.10}$$

where the first two terms in (2.10) are given by (2.4) and (2.5), respectively, and

$$V'' = \sum_{1=i<j<k} V(r_i, r_j, r_k) \tag{2.11}$$

is the three-body potential.

The matrix elements of the residual two-body interaction $V'$ (equation (2.5)) with respect to the bases (equation (2.6)) are given in detail in [8,10,28,33] by using two-particle FPCs; specifically, they are products of orbital and spin-isospin two-particle FPCs. Similarly, the matrix elements of the three-body interaction are calculated by using three-particle FPCs. The ground- and excited-state nuclear wave functions, which are obtained via the diagonalization of the energy matrices, are used to calculate the root-mean-square radius, the magnetic dipole moment and the total integral cross section for the photoabsorption of $\gamma$ quanta by the nucleus.

## 3. The Root-Mean-Square Radius and the Magnetic Dipole Moment

The root-mean-square radius is defined as

$$\mathcal{R} = \sqrt{r_p^2 + \langle R_{Nuc}^2 \rangle}, \tag{3.1}$$

where $r_p = 0.85$ fm is the proton radius and the second term in the sum is the mean value of the following operator [28]:

$$R_{Nuc}^2 = \frac{1}{A^2} \sum_{1=i<j}^{A} r_{ij}^2. \tag{3.2}$$

This operator does not depend on the spin-isospin variables of the nuclear wave function, and its calculation is straightforward [28].

The nuclear magnetic dipole moment is defined as the mean value of the operator $\hat{\mu} = \hat{\mu}_\sigma + \hat{\mu}_0$, where [2]

$$\hat{\mu}_\sigma = \sum_{i=1}^{A} [(\mu_p + \mu_n) + 2(\mu_p - \mu_n) t_{0i}] s_{0i} \tag{3.3}$$

and

$$\hat{\mu}_0 = \frac{1}{2} \sum_{i=1}^{A} [(1 - 2t_{0i})] \ell_{0i}, \tag{3.4}$$

calculated in a state with $M_J = J$. In equations (3.3) and (3.4), $\mu_p$ and $\mu_n$ are the proton and neutron magnetic moments, respectively, and $t_{0i}$, $s_{0i}$ and $\ell_{0i}$ are the z components of the isospin, spin and orbital momenta, respectively, of the $i$th nucleon. By writing each of the two operators $\hat{\mu}_\sigma$ and $\hat{\mu}_0$ as a sum of symmetric and antisymmetric operators of symmetry types $[A]$ and $[A-1,1]$, in the forms

$$\hat{\mu}_\sigma = \mu_\sigma^{[A]} + \mu_\sigma^{[A-1,1]}, \quad \hat{\mu}_0 = \mu_0^{[A]} + \mu_0^{[A-1,1]}, \tag{3.5}$$

the mean value of the magnetic dipole moment can be transformed into an algebraic expression that depends on the orbital and spin-isospin quantum numbers of the $A$-nucleon state, and the calculations are then straightforward [2].

## 4. The Partial Integral Cross Section of $\gamma$-Quanta Photoabsorption

The partial integral cross section of the dipole electric transition that accompanies $\gamma$-quanta photoabsorption is calculated by using the well-known line integral [6,34]

$$\sigma_{i \to f} = \int \sigma dE_f = (2\pi)^3 \frac{e^2}{\hbar c} \frac{(\hbar c)^2}{E_\gamma} \sum_{\mu M_j'} |\langle f : J'M_J'T'M_T'|T_{1\mu}^{elec}|i : JM_JTM_T\rangle|^2, \qquad (4.1)$$

where $E_\gamma = E_f - E_i$ is the energy of the $\gamma$ quanta and $|i\rangle$ and $|f\rangle$ are the wave functions of the initial and final states, respectively. The operator for the dipole electric $\gamma$ transition, $T_{1\mu}^{elec}$, is defined as [6,34]

$$T_{1\mu}^{elec} = \frac{\sqrt{2k}}{3} \sum_{i=1}^{A} t_0(i) r(i) y_{1\mu}(i) = \frac{\sqrt{2k}}{3} \frac{1}{A-1} \sum_{1=i<j}^{A} \sqrt{\frac{3}{4\pi}} [t_0(i) r_{1\mu}(i) + t_0(j) r_{1\mu}(j)], \qquad (4.2)$$

where $k$ is the angular wave number, $k = \frac{E_\gamma}{\hbar c}$. In equation (4.2), we have transformed the Cartesian components of the vectors into their corresponding spherical components, in the usual manner, as follows:

$$A_\mu = \sqrt{\frac{4\pi}{3}} A Y_{1\mu}(\hat{A}), \mu = 0, 1, -1. \qquad (4.3)$$

If one-particle FPCs are not allowed for higher configuration spaces, i.e., higher values of $N$, one can use the following transformations:
(1) Introduce the centre-of-mass coordinates and the relative coordinates in the forms

$$\boldsymbol{R}_{ij} = \frac{1}{2}(\boldsymbol{r}_i + \boldsymbol{r}_j) \text{ and } \boldsymbol{r}_{ij} = \boldsymbol{r}_i - \boldsymbol{r}_j. \qquad (4.4)$$

(2) Use similar relations for the isotopic spin components:

$$T_0(ij) = t_0(i) + t_0(j), t_0(ij) = \frac{1}{2}\{t_0(i) - t_0(j)\}. \qquad (4.5)$$

By applying the graphical method for the addition of angular momenta [34], one can calculate $\sigma_{i \to f}$. The total integral cross section of the dipole electric transition accompanying the photoabsorption of $\gamma$ quanta by $^4$He, $\sigma$, is then the sum of the $\sigma_{i \to f}$ for all possible final states $|f\rangle$.

The methods of calculating the one-, two-, three- and four-particle FPCs in the USM are given in [35]. In addition, recurrence relations for the two-particle orbital FPCs and tables of these coefficients for $3 \leq A \leq 6$ and $N \leq 3$ are given by Vanagas [2]. A general and direct method for calculating the two-particle orbital FPCs and tables of these coefficients for $A = 6$ and $2 \leq N \leq 4$ are given by Doma and Machabeli [36]. Furthermore, this direct method has been used to calculate the two-particle orbital FPCs for nuclei with $A = 3$ and $0 \leq N \leq 10$ in [37]. Finally, in the present study, we calculated the necessary orbital FPCs for nuclei with $A = 3$ and $4$.

## 5. Results and Discussions

In our investigation, the ground- and excited-state wave functions of $^3$H and $^4$He were expanded in series in terms of the basis functions of the USM. For $^3$H, bases corresponding to even numbers of quanta of excitation $N$ in the range $0 \leq N \leq 20$ were constructed. For $^4$He, bases corresponding to both even and odd numbers of quanta of excitation $N$ in the range $0 \leq N \leq 20$ were constructed. Each of these basis functions was expanded in terms of one-, two-and three-particle total FPCs, which are products of orbital and spin-isospin coefficients. As a result, the Hamiltonian matrices for the different states of $^3$H and $^4$He were constructed as functions of the oscillator parameter $\hbar\omega$. By diagonalizing these matrices with respect to $\hbar\omega$ as a variational parameter, we obtained the nuclear energy eigenvalues and the corresponding eigenfunctions. Furthermore, the matrix elements of the different operators corresponding to different nuclear characteristics were calculated.

The ground state of triton has a total angular momentum of $J = \frac{1}{2}$, an isotopic spin of $T = \frac{1}{2}$, and even parity, i.e., $(J^\pi, T) = \left(\frac{1}{2}^+, \frac{1}{2}\right)$. The energy eigenvalues obtained via the diagonalization of the Hamiltonian matrices for the state $\left(\frac{1}{2}^+, \frac{1}{2}\right)$ of triton for each value of the oscillator parameter $\hbar\omega$, which was allowed to vary in the range $8 \leq \hbar\omega \leq 20$ MeV, showed two accepted values: the lower one belongs to the ground state, and hence, the higher and negative binding energy value corresponds to the first-excited-state energy $E^*$. Other higher values were also obtained, but we do not present them here because there is no experimental evidence for the existence of these excited states of triton. The obtained ground-state wave functions were used to calculate the root-mean-square radius and magnetic dipole moment of triton.

In Table-1, we present various quantities that characterize the ground-state wave function of triton. For this purpose, we present the $S$-, $D$- and $P$-state probabilities, denoted by $P_S$, $P_D$ and $P_P$, respectively, for the different potentials considered. The probabilities of the bases having irreducible representations $[f] = [3], [21]$ and $[111]$ of the symmetry group $S_3$ for the ground-state wave functions of triton are also given in this table. Obviously, $P_{S=\frac{1}{2}} = 1 - P_D$ and $P_{S=\frac{3}{2}} = P_D$. Previous results [22] obtained by using the Faddeev equation with the Argonne $v18$ nucleon-nucleon interaction plus the UIX three-nucleon interaction are shown. Furthermore, previous results [38] obtained by using the technique of hyperspherical harmonics [39] with the effective AV18 nucleon-nucleon potential plus the UIX three-nucleon interaction are also shown, along with previous results [40] obtained by using the *ab initio* few-body method with the effective AV8' + 3NF potentials.

Table-1 The $S$-, $D$- and $P$-state probabilities ($P_S$, $P_D$ and $P_P$, respectively) for the ground-state wave function of triton and the probabilities of the bases having irreducible representations $[f] = [3], [21]$ and $[111]$ of the symmetry group $S_3$. Previous results [22] obtained by using the Faddeev equation with the Argonne $v18$ and UIX three-nucleon interactions are shown. In addition, previous results [38] obtained by using the hyperspherical harmonics method with the effective AV18 + UIX potentials are also shown, along with previous results [40] obtained by using the *ab initio* few-body method with the effective AV8' + 3NF potentials.

| Case Charact. | GPT | GPT + UIX | AV8' | AV8' + UIX | Faddeev, AV18+ UIX [22] | Hyperspherical harmonics AV18 + UIX [38] | *ab initio* AV8' + 3NF [40] |
|---|---|---|---|---|---|---|---|
| $P_S$% | 92.84 | 91.856 | 91.894 | 91.204 | 90.563 | 90.565 | -- |
| $P_D$% | 6.73 | 7.907 | 7.877 | 8.448 | 9.302 | 9.300 | 8.69 |
| $P_P$% | 0.43 | 0.237 | 0.229 | 0.348 | 0.135 | 0.135 | -- |
| $P_{[3]}$% | 86.38 | 88.724 | 86.774 | 86.251 | -- | -- | -- |
| $P_{[21]}$% | 13.56 | 11.224 | 13.178 | 13.603 | -- | -- | -- |
| $P_{[111]}$% | 0.06 | 0.052 | 0.048 | 0.146 | -- | -- | -- |

In Table-2, we present for triton the binding energy in MeV, the root-mean-square radius in fm, the first-excited-state energy in MeV, and the magnetic dipole moment in N.M. calculated by using the two nucleon-nucleon potentials considered in this study (GPT and AV8'). The improved values that resulted from including the UIX three-nucleon interaction are also given. Moreover, the corresponding experimental values and the values of the oscillator parameter $\hbar\omega$ that produced the minimum ground-state energy eigenvalues are provided. Previous results obtained by using the Faddeev equation together with the AV18 nucleon-nucleon interaction plus the UIX three-body interaction are shown [22]. Previous results [12] obtained by using the NCSM with the AV18 nucleon-nucleon interaction plus the TM three-body interaction are also shown, along with results obtained by using the *ab initio* few-body method with the effective AV8' + 3NF potentials.

Concerning the second $\left(\frac{1}{2}^+, \frac{1}{2}\right)$ state of $^3$H, we note that the three- and four-nucleon states each have only one bound state, and all excited states are in the continuum. The use of a bound-state approach with square-integrable basis functions is only meaningful for bound states and narrow resonances. In contrast to $^4$He, which exhibits a broad $0^+$ resonance, there are no resonances in the $\frac{1}{2}^+$ channel of tritium. The reason why its energy increases with increasing HO frequency and does not exhibit a clear minimum is that this eigenstate and those above it represent a discretization of the energy continuum, and as such, they continuously move as the size of the model space is increased or as other parameters are varied.

Table-2 For $^3$H, the binding energy (B.E.) in MeV, the first-excited-state energy ($E^*$) in MeV, the root-mean-square radius ($R$) in fm, and the magnetic dipole moment ($\mu$) in N.M. calculated by using the GPT and AV8' nucleon-nucleon interactions as well as the improved results obtained by using the UIX three-nucleon interaction. The values of $\hbar\omega$ that produced the best fit to the binding energy of $^3$H are also given. Previous results obtained by using the Faddeev equation with the AV18 nucleon-nucleon interaction plus the UIX three-nucleon interaction are shown [22], along with previous results [12] obtained by using the NCSM with the AV18 nucleon-nucleon interaction plus the TM three-body interaction. The experimental values are also provided. Moreover, results obtained by using the *ab initio* few-body method with the effective AV8' + 3NF potentials are shown.

|  | GPT | GPT + UIX | AV8' | AV8' + UIX | Faddeev AV18 + UIX [22] | NCSM AV18 + TM [12] | *ab initio* AV8' + 3NF [40] | Exper. [41] |
|---|---|---|---|---|---|---|---|---|
| B.E. | 8.2816 | 8.4283 | 8.3081 | 8.4753 | 8.470 | 8.33 | 8.41 | 8.482 |
| $E^*$ | 8.6034 | 8.1661 | 8.5086 | 8.1315 | --- | 6.2065 | --- | --- |
| $R$ | 1.8163 | 1.7570 | 1.7989 | 1.7521 | --- | --- | 1.70 | 1.75 |
| $\mu$ | 3.362 | 3.2455 | 3.2865 | 3.2311 | --- | --- | -- | 2.979 |
| $\hbar\omega$ | 13 | 13 | 14 | 14 | --- | --- | -- | --- |

The ground state of $^4$He is $(0^+, 0)_1$, and its excited states are $(0^+, 0)_2$, $(0^-, 0)$, $(2^-, 0)$, $(2^-, 1)$, $(1^-, 1)_1$, $(0^-, 1)$, $(1^-, 1)_2$, $(2^+, 0)$ and $(1^-, 0)$. Another excited state, $(1^+, 0)$, was predicted for this nucleus in [28]. The Hamiltonian matrices that belong to these states were constructed with respect to numbers of quanta of excitation $N$ in the range $0 \leq N \leq 20$ as functions of the oscillator parameter $\hbar\omega$, which was allowed to vary over a large range of values, $8 \leq \hbar\omega \leq 28$ MeV, to obtain the best fit to the spectrum of $^4$He. Among all levels of $^4$He, the dipole transition is allowed only for levels with $J = 1$ and $T = 1$, i.e., the levels $(1^-, 1)_1$ and $(1^-, 1)_2$. The eigenvalues that resulted from the diagonalization of the ground-state Hamiltonian matrices for $^4$He in this study showed two accepted values: the lower one belongs to the ground state, $(0^+, 0)_1$, and the higher belongs to the first excited state, $(0^+, 0)_2$. The obtained ground- and excited-state nuclear wave functions were used to calculate the spectrum, the root-mean-square radius and the total integral cross section of the dipole electric transition accompanying $\gamma$-quanta photoabsorption for the $^4$He nucleus with respect to each value of the oscillator parameter $\hbar\omega$.

In Table-3, we present different quantities that characterize the ground-state wave function of $^4$He. In this table, we present the probabilities $P_S$, $P_D$ and $P_P$ for the ground-state wave function of $^4$He obtained by using the two nucleon-nucleon interactions alone as well as the improved values resulting from the inclusion of the three-nucleon interaction. The probabilities of bases having irreducible representations $[f] = [4], [31], [22]$ and $[211]$ of the symmetric group $S_4$ for the ground-state wave functions of $^4$He are also shown, as are previous results obtained by using the NCSM [42].

In Table-4, we present the results obtained for $^4$He by using the GPT and AV8' nucleon-nucleon interactions to calculate the binding energy in MeV, the root-mean-square radius in fm, and the total integral cross section of the dipole electric transition accompanying the photoabsorption of $\gamma$-quanta by this nucleus ($\sigma$) in MeV·mbarn. The improved values obtained by including the UIV three-nucleon interaction are also given. Moreover, the corresponding experimental values and the values of the oscillator parameter $\hbar\omega$ that best reproduced the spectrum of $^4$He are also given in Table-4. Previous results obtained by using the Faddeev-Yakubovsky method [25] and the NCSM method [25] are also shown.

In Fig. 1, we present the spectra of $^4$He that resulted from using the two nucleon-nucleon interactions alone and from including the three-nucleon interaction. The experimental spectrum [45] is also shown in this figure. It is seen that the order of the calculated levels is correct and that the calculated spectra are in good agreement with the corresponding experimental data for both considered interactions. Interestingly, a new even-parity level for $^4$He is obtained near the threshold energy value. This is the level $(1^+, 0)$, with an energy equal to 32.61, 31.14, 33.39 or 32.10 MeV according to the GPT, GPT + UIX, AV8' and AV8' + UIX potentials, respectively. The main contributions to this level are due to bases with $[f] = [31]$, whereas bases with $[f] = [221]$ contribute little.

Table-3 The probabilities $P_S$, $P_D$ and $P_P$ for the ground-state wave function of $^4$He for the different considered interactions. The probabilities of bases having irreducible representations $[f] = [4], [31], [22]$ and $[211]$ of the symmetric group $S_4$ are also given. Obviously, $P_{S=0} = P_S$, $P_{S=2} = P_D$, and $P_{S=1} = P_P$. Previous results obtained by using the NCSM are also shown.

| Case Characteristic | GPT | GPT + UIX | AV8' | AV8' + UIX | NCSM [42] |
|---|---|---|---|---|---|
| $P_S\%$ | 91.570 | 88.856 | 89.946 | 89.388 | 86.73 |
| $P_D\%$ | 6.194 | 9.143 | 8.832 | 9.222 | 12.98 |
| $P_P\%$ | 2.236 | 2.001 | 1.222 | 1.390 | 0.29 |
| $P_{[4]}\%$ | 91.540 | 88.684 | 90.113 | 89.674 | -- |
| $P_{[31]}\%$ | 1.732 | 2.522 | 1.962 | 1.576 | -- |
| $P_{[22]}\%$ | 6.224 | 8.481 | 7.683 | 8.459 | -- |
| $P_{[211]}\%$ | 0.504 | 0.313 | 0.242 | 0.291 | -- |

Table-4 For $^4$He, the binding energy (B.E.) in MeV, the root-mean-square radius ($R$) in fm, and the calculated value of the total integral cross section of $\gamma$-quanta photoabsorption ($\sigma$) in MeV·mbarn obtained by using the GPT and AV8$'$ nucleon-nucleon interactions as well as the improved values obtained by including the UIX three-nucleon interaction. The corresponding experimental values and the values of the oscillator parameter $\hbar\omega$ (in MeV) that resulted in the best fit between the calculated $^4$He spectra and the experimental values are also given. Previous results obtained by using the Faddeev-Yakubovsky (FY) method [25] and the NCSM method [25] are also shown in this table.

|  | GPT | GPT + UVII | AV8$'$ | AV8$'$ + UIX | FY [25] | NCSM [25] | Exper. |
|---|---|---|---|---|---|---|---|
| B.E. | 27.929 | 28.272 | 27.999 | 28.279 | 25.94 | 25.80 | 28.296 [43] |
| $R$ | 1.670 | 1.520 | 1.666 | 1.511 | 1.485 | 1.485 | 1.46 [44] |
| $\sigma$ | 70.48 | 68.995 | 69.464 | 69.321 | -- | -- | -- |
| $\hbar\omega$ | 17 | 17 | 16 | 16 | -- | -- | -- |

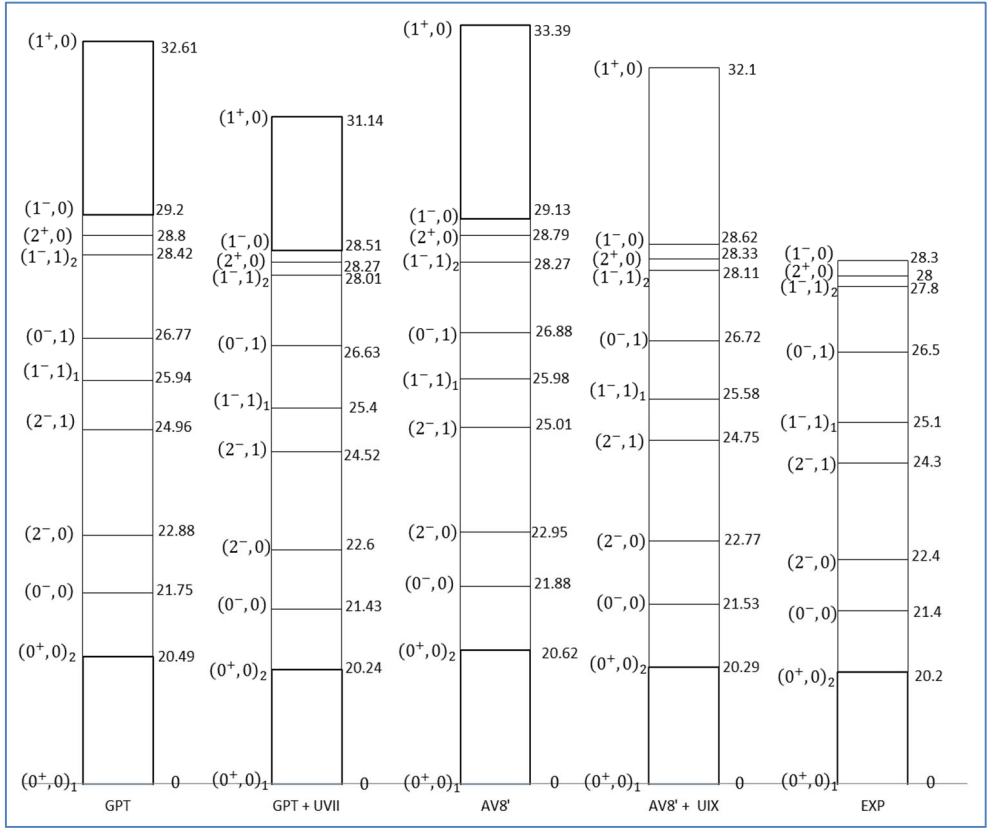

Fig. 1 The $^4$He spectra calculated by using the GPT, GPT + UVII, AV8$'$, and AV8$'$ + UIX potentials along with the experimental spectrum [45].

Table-2 and Table-4 show that the inclusion of the three-body interaction improved the calculated ground- and excited-state characteristics of $^3$H and $^4$He for both nucleon-nucleon interactions, as expected. It is also seen from Table-2, Table-4 and Fig. 1 that the values of the different characteristics of $^3$H and $^4$He calculated by using the AV8$'$ and GPT potentials alone and by including the UIX three-body interaction are all in good agreement with the corresponding experimental values. Furthermore, Table-4

shows that the inclusion of the three-nucleon interaction reduced the calculated value of the total integral cross section of the photoabsorption of $\gamma$-quanta by the $^4$He nucleus ($\sigma$) and that both potentials yielded approximately the same values of $\sigma$, with the absorption lying in the energy range from 25.40 to 28.27 MeV.

Moreover, to study the convergence properties of the USM approach with respect to the dimensionality of the adopted model space, the results of the USM calculations for $N \leq 20$ were incrementally extrapolated to $N = 30$ by using Stirling's formula [46] for both nuclei. In Figs. 2, 3, 4 and 5, we present the convergence behaviours of the binding energy, the root-mean-square radius, the first-excited-state energy and the magnetic dipole moment, respectively, for triton with respect to the number of quanta of excitation $N$. In Figs. 6, 7 and 8, we present the convergence behaviours of the binding energy, the root-mean-square radius and the first-excited-state energy, respectively, for $^4$He with respect to the number of quanta of excitation $N$. These figures show that the calculated characteristics of $^3$H and $^4$He converge to the corresponding experimental values as the value of $N$ increases.

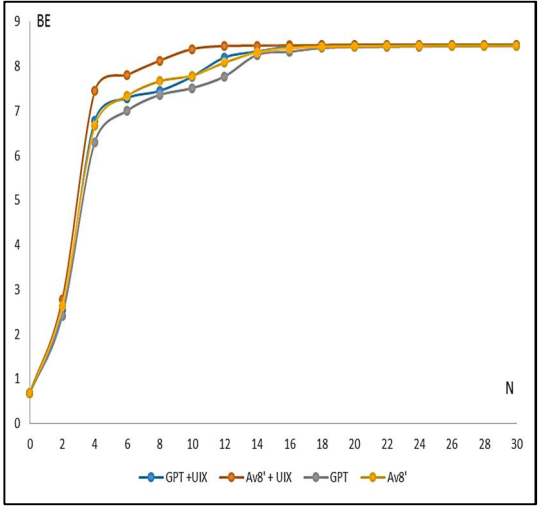
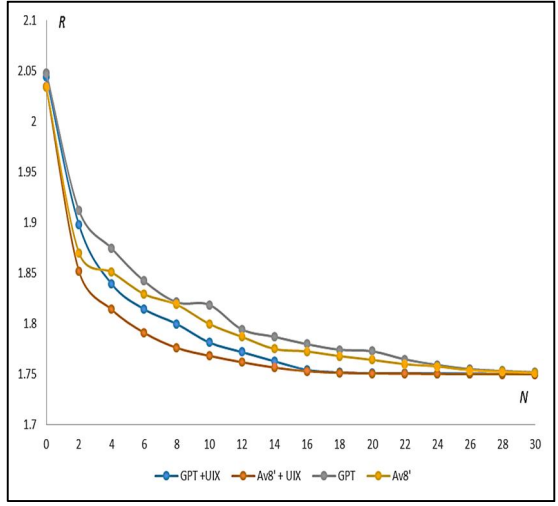

Fig. 2                                                  Fig. 3

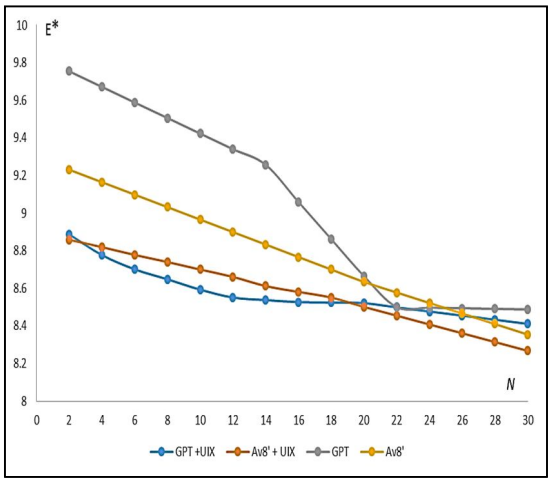
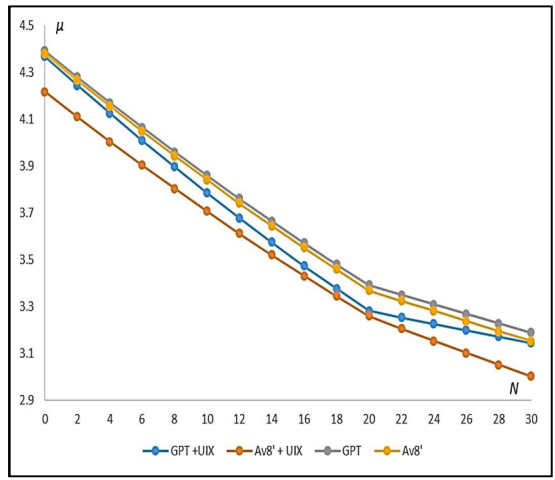

Fig. 4                                                  Fig. 5

Figs. 2, 3, 4 and 5 Convergence behaviour of the $^3$H characteristics.

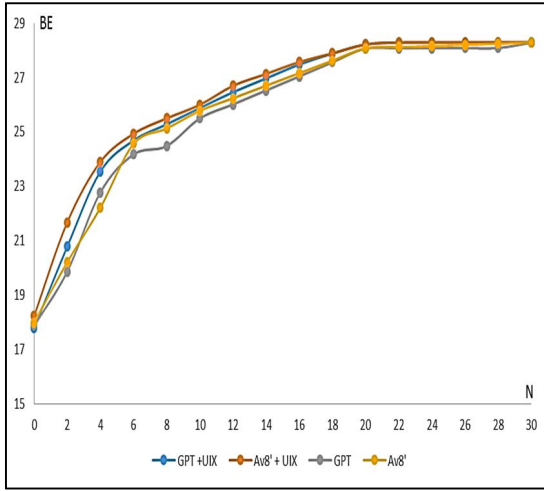
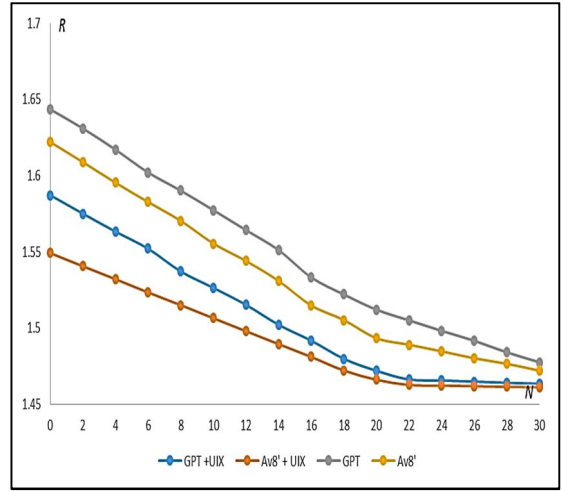

Fig. 6            Fig. 7

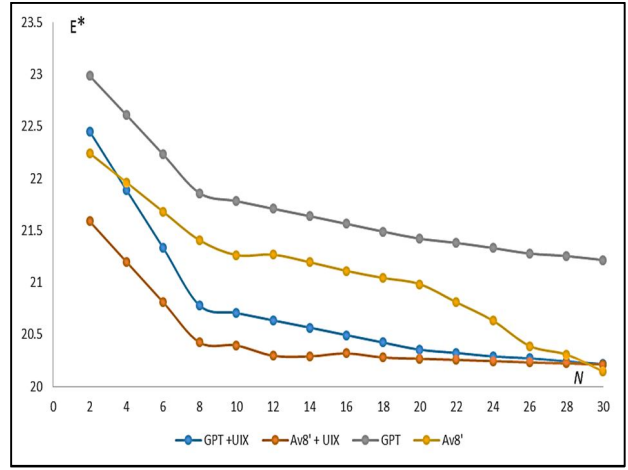

Fig. 8

Figs. 6, 7 and 8 Convergence behaviour of the $^4$He characteristics.

## 6. Conclusion

The results obtained in this study show that three-body interactions play a very important role in the calculation of the different ground- and excited-state characteristics of $^3$H and $^4$He. Moreover, high configurations, with suitably chosen bases, also play a major part in determining the convergence behaviours of the nuclear characteristics. Calculations within a three-body translationally invariant HO basis using realistic two- and three-nucleon forces have previously been performed for the three-nucleon system (see, e.g., [47]) and for the spectrum and root-mean-square radius of the four-nucleon system (see, e.g., [48,49]). These previous calculations were based on the translationally invariant form of the no-core shell model, which is completely equivalent to the USM except that the antisymmetrization of the wave function is achieved not by means of group theory but rather by diagonalizing the antisymmetrization operator and retaining the antisymmetric eigenstates as basis states. Furthermore, the spectrum and root-mean-square radius of the three- and four-nucleon systems have been studied by means of numerically exact few-body approaches (such as the Faddeev, Faddeev-Yakubovsky and hyperspherical harmonics approaches starting from realistic two- and three-nucleon forces; see, for example, [22, 23, 25]). A review can be found in [24]. By contrast, the calculations presented here, in addition to providing a complete picture of the nuclear structure based on the obtained nuclear wave functions of the different states of the investigated nuclei, also present a certain degree of novelty with respect to the translationally invariant no-core shell model. In particular, the direct construction of the antisymmetric three- and four-body basis states by means of group theory is more elegant and may even prove to be computationally more advantageous, considering that all calculations of the nuclear characteristics reported in the present paper were performed by applying group-theoretical methods.